\documentclass[twoside]{article}
\usepackage{fleqn,espcrc2}
\usepackage{epsfig}
\usepackage{amssymb}

\title{Hadron masses from dynamical, non-perturbatively $O(a)$
improved Wilson fermions}

\author{%
H.~St\"uben%
\address{Konrad-Zuse-Zentrum f\"ur Informationstechnik Berlin, 
14195 Berlin, Germany},
\emph{QCDSF} and \emph{UKQCD} Collaborations}

\newcommand{\mPS}{m_{\mathrm{PS}}}
\newcommand{\mV}{m_{\mathrm{V}}}
\newcommand{\MV}{M_{\mathrm{V}}}
\newcommand{\mN}{m_{\mathrm{N}}}

\begin{document}

\begin{abstract}
We present results on light hadron masses from simulations of full QCD
and report on experiences in running such simulations on a Hitachi
SR8000-F1 supercomputer.
\end{abstract}

\maketitle

\section{INTRODUCTION}

With the installation of a 112 node Hitachi SR8000-F1 at the
Leibniz-Rechenzentrum in Munich the QCDSF Collaboration has started
simulating QCD with two degenerate flavors of fermions.  In a
cooperative program with the UKQCD Collaboration (see talk by Irving
\cite{Irving}) we share configurations giving both collaborations the
possibility to do measurements at a larger set of bare parameters.

This contribution has two parts.  In the first part we report on
performance of QCD calculations on the Hitachi SR8000-F1.  In the
second part we present results on the light hadron mass spectrum
for dynamical, non-perturbatively $O(a)$ improved Wilson fermions.  We
focus on comparison with our quenched results \cite{QCDSF,Pleiter}.

\section{QCD ON THE HITACHI SR8000-F1}

\subsection{Hardware overview}

The SR8000 in Munich is a ``cluster'' of 112 symmetric multiprocessor
nodes.  Each node has 8 CPUs and (at least) 8 GByte of shared memory.
Each CPU has a a 128 kByte 4-way associative cache, 160 registers and
has a pseudo-vectorization facility.  The peak performance of a CPU is
1.5~Gflop/s giving a peak performance of 1344 Gflop/s for the whole
machine.

\subsection{Pseudo-vectorization}

It order to get good performance it is essential to employ
pseudo-vectorization.  At the hardware level pseudo-vectorization is
implemented by efficient pre-fetch and pre-load mechanisms.  At the
programming level the compiler tries to vectorize most inner loops.
These loops are vectorized under the usual conditions, i.e., if there
are no data access conflicts, no function/subroutine calls and no
if-blocks in the loop body.

On the SR8000 one can basically use the same code as on
a standard RISC processor and more important one can have the same data
layout as for a RISC CPU.  For example our data layout is (in
Fortran~90 notation)
\begin{verbatim}
  complex(8) u(3, 3, vol/2, even:odd, dim)
  complex(8) a(4, 3, vol/2)
  complex(8) c(18, 2, vol/2, even:odd)
\end{verbatim}
for the gauge, pseudo-fermion and clover field.  In contrast to a real
vector computer the vector index does not have to be shifted to the
first place.

\begin{figure*}[!t]
\epsfig{file=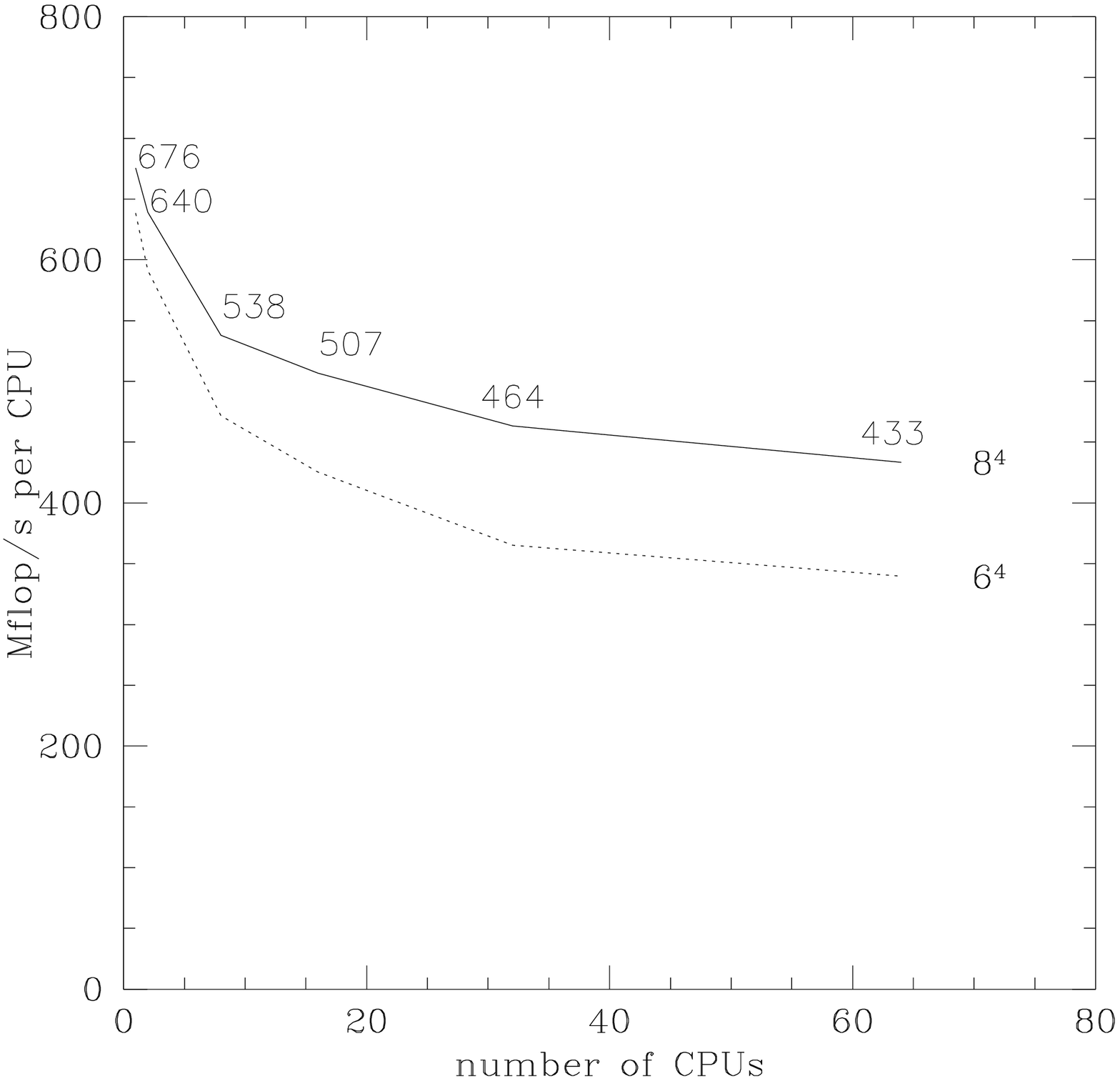,width=7.5cm,bbllx=0,bblly=0,bburx=568,bbury=540}%
\hspace{1cm}%
\epsfig{file=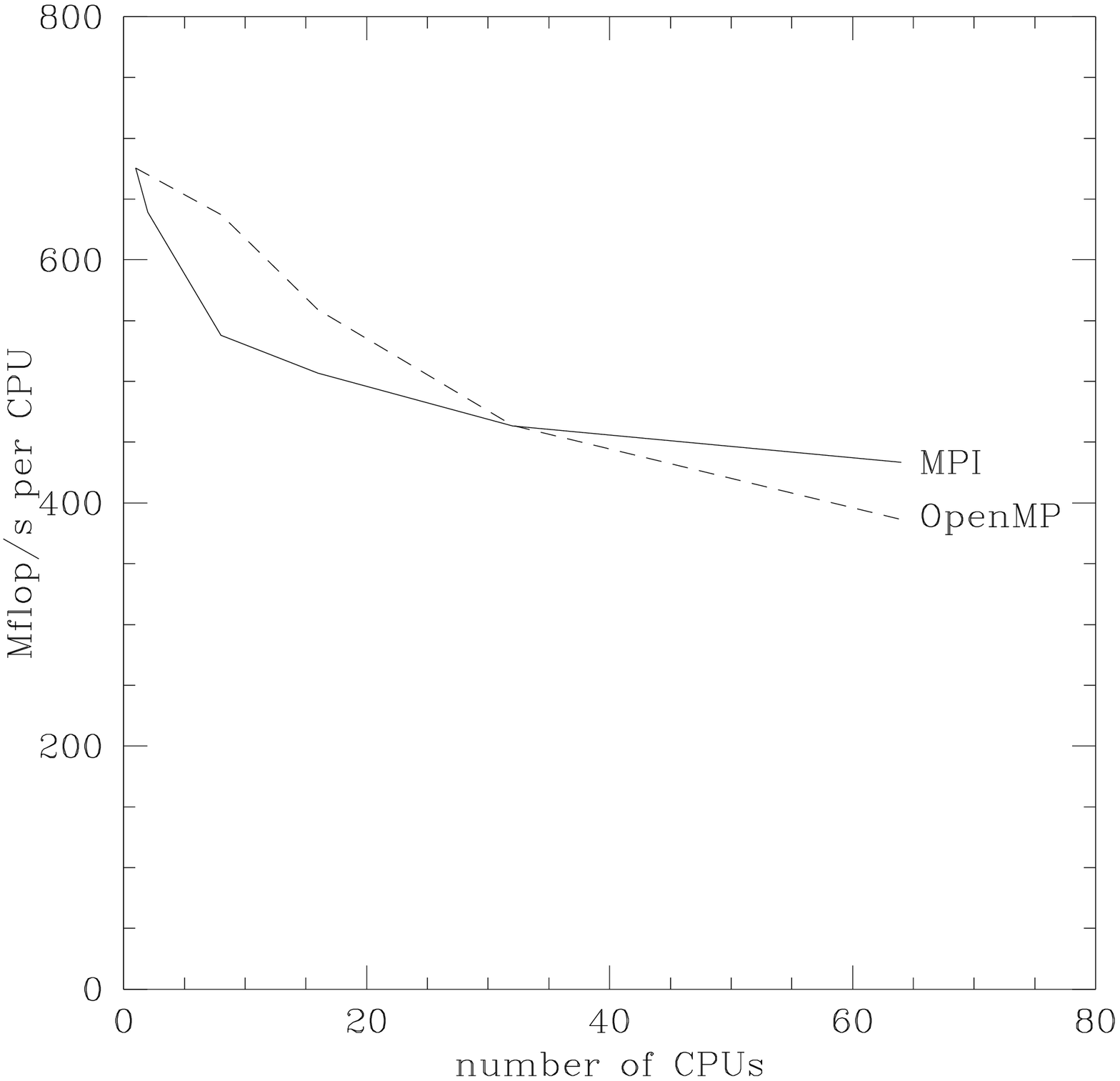,width=7.5cm,bbllx=0,bblly=0,bburx=568,bbury=540}
\vspace{-9mm}
\caption{Communication loss for constant lattice volume per CPU as
indicated.  Further explanations are given in the text.}
\end{figure*}

\vspace*{-0.2\baselineskip}

\subsection{Single CPU performance}

To fix the notation let $M$ be the fermion matrix, $D$ the Wilson
hopping term and $A$ the clover term:
\begin{equation}
M = A - \kappa D
\end{equation}
\begin{equation}
A = 1 - \kappa c_{\mathrm{sw}} \sigma_{\mu\nu} \cal{F}_{\mu\nu}
\end{equation}
Performance figures for the multiplication of $D$, $A$, $A^{-1}$ and
$M$ with a vector are given in Table~1.  These figures were obtained
with Fortran~90 code using 64 bit arithmetic on an $8^4$ lattice.  The
code performs well in comparison with RISC processors where one
typically sustains 10--20~\% of the peak performance.

\begin{table}[htb]
\caption{Performance of the multiplication with a vector on a single
CPU.}
\begin{tabular}{crr}
\hline
$D$ &      640 Mflop/s & (43 \% of peak)\\
$A$ &     1160 Mflop/s & (77 \% of peak)\\
$A^{-1}$ & 630 Mflop/s & (42 \% of peak)\\[\medskipamount]
$M$ &      676 Mflop/s & (45 \% of peak)\\
\hline
\end{tabular}
\end{table}

\vspace*{-0.2\baselineskip}

\subsection{Parallelization}

Our code is parallelized with MPI.  In addition we have experimented
with a hybrid programming model using OpenMP for distributing the work
over the CPUs of a single node and MPI for the communication between
nodes.  Figure~1 shows the communication loss for the fermion matrix
multiplication.  

The solid lines (being identical in both plots) show performance
measurements for a lattice volume of $8^4$ per CPU for the pure MPI
code.  In the left plot these are compared with measurements on
volumes of $6^4$ per CPU.  In the right plot comparison is done with
the hybrid programming model again on a CPU local volume of $8^4$.

As can be seen from Figure~1 the communication loss is quite
substantial.  The observations on the pure MPI code can be understood
qualitatively in the following way.  Using more and more CPUs the
communication pattern becomes more and more complex.  
Within one node when using 2, 4 or 8 CPUs there are 2, 4 or 6
boundaries which have to be communicated and the surface/volume
ratio is 2/8, 4/8 or 6/8 respectively.  

Between nodes communication is along 1, 2
or 3 dimensions (in our program the lattice can be decomposed
maximally in three dimensions) when using 2, 4, or 8 (and more) nodes
leading again to more and more complicated communication patterns.

On a single node parallelization with OpenMP is more efficient than
with MPI.  This is to be expected because there is no communication
needed in that case.  But surprisingly this gain does not persist when
switching on inter node communication.  In both cases the same amount
of data has to be transferred between the nodes.  But in the pure MPI
version more calls are needed to copy data between nodes and in
addition data has to be copied within the nodes.  Both should lead to
larger communication overhead.

According to the hardware performance monitor we are running at
about 400--450 Mflop/s per CPU in our production runs.  This is in
accordance with the 433 Mflop/s given in Figure~1.

\section{LIGHT HADRON MASSES}

\subsection{Simulation details}

We investigate full QCD with the standard Wilson action for the
plaquette and the fermions plus the $O(a)$ improvement term.  For the
improvement coefficient $c_{\mathrm{sw}}(\beta)$ we use the
non-perturbative values determined by the ALPHA collaboration
\cite{ALPHA}.

Simulation parameters and statistics are listed in Table~2. In these
simulations the $\mPS/\mV$ ratio lies between 0.60 and 0.83.  The
Sommer scale \cite{Sommer} $r_0/a$ varies between 4.7 and 5.5
\cite{Irving}.

\vspace{-1\baselineskip}

\begin{table}[h]
\caption{\vspace*{-\baselineskip}}
\begin{tabular}{cccr}
\hline
$\beta$ & $\kappa_{\mathrm{sea}}$ & $V$ & $N_{\mathrm{traj}}$ \\
\hline
\multicolumn{4}{c}{UKQCD} \\
\hline
5.20  & 0.1350 & $16^3\times32$ & 6000 \\
      & 0.1355 & $16^3\times32$ & 8000 \\[\medskipamount]

5.26  & 0.1345 & $16^3\times32$ & 4000 \\[\medskipamount]

5.29  & 0.1340 & $16^3\times32$ & 4000 \\
\hline
\multicolumn{4}{c}{QCDSF} \\
\hline
5.29  & 0.1350 & $16^3\times32$ & 2270 \\
      & 0.1355 & $24^3\times48$ &  605 \\
\hline
\end{tabular}
\end{table}

\vspace{-1\baselineskip}

Figure~2 shows the integrated autocorrelation time of the plaquette
$\tau_{\mathrm{int},P}(t) = \frac{1}{2} + \sum_{t'=1}^{t} \rho_P(t')$
($\rho_P$ is the normalized autocorrelation function)
for the QCDSF run on the $16^3\times32$ lattice.  From the plot
we read off that in this case $\tau_{\mathrm{int},P} \lesssim 10$.

\subsection{Results}

Figure~3 shows an APE plot.  For clarity error bars for the quenched
results (0.1--3.5~\%) are not shown.  One sees that for currently reachable sea
quark masses unquenching has, if any, only small effects on the light
hadron spectrum and that with dynamical fermions it is a long
way to $\mPS/\mV$ ratios as small as in the quenched case.

In the following we compare our dynamical results with chiral
extrapolations done in the quenched case at $\beta = 6.0$ where $r_0/a
= 5.36$.  This is within the range of $r_0/a$ of the dynamical runs
studied.  Let us first recall the chiral extrapolation of $\mV$.  Our
quenched results for $\mV$ are very well described by
a phenomenological fit
\begin{equation}
(a\mV)^2 = (a\MV)^2 + b_2 (a\mPS)^2 + b_3 (a\mPS)^3
\end{equation}

\pagebreak

\vspace*{-0.5\baselineskip}

\begin{figure}[t]
\epsfig{file=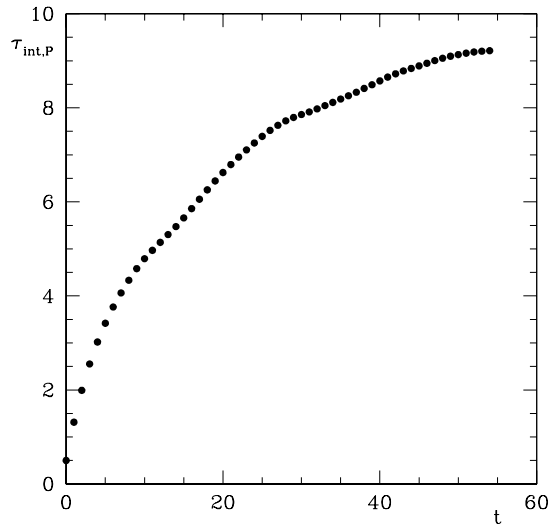,width=7.5cm,bbllx=0,bblly=0,bburx=568,bbury=540}
\vspace{-14.5mm}
\caption{Integrated autocorrelation time of the plaquette for the run
at $\beta=5.29$ and $\kappa=0.1350$.}
\end{figure}

\vspace*{-4\baselineskip}
\begin{figure}[h]
\epsfig{file=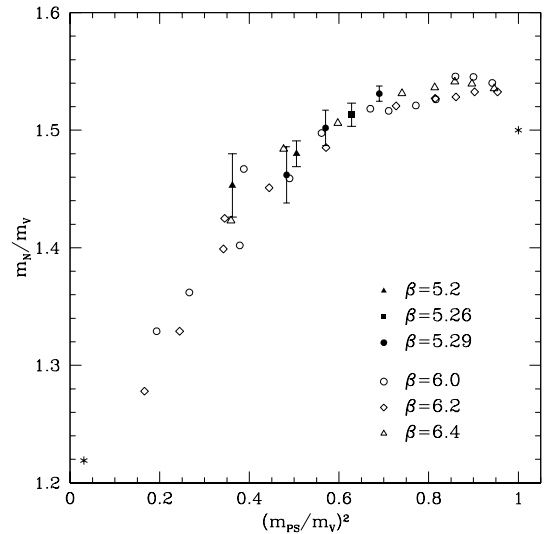,width=7.5cm,bbllx=0,bblly=0,bburx=568,bbury=540}
\vspace{-14.5mm}
\caption{APE plot. Open symbols represent our quenched results
\cite{QCDSF,Pleiter} without error bars.  Filled symbols are our
dynamical results.}
\end{figure}
\vspace*{-0.5\baselineskip}

\noindent
where $b_2 = 0.910(20)$ and $b_3 = 0.049(15) $\cite{Pleiter}.  The
lightest three points were excluded from the fit.  The deviation of
these points is believed to be a quenching artifact
\cite{QCDSF,Pleiter}.  A quenching artifact appearing in quenched
chiral perturbation theory for the vector meson mass is a term
$\propto a\mPS$ \cite{Booth}.

\pagebreak

\begin{figure}[t]
\epsfig{file=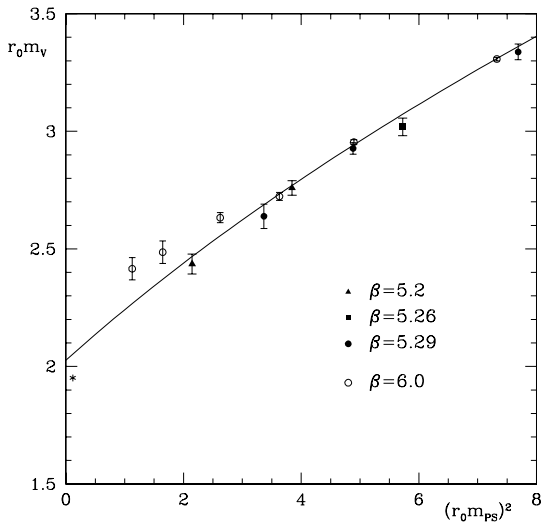,width=7.5cm,bbllx=0,bblly=0,bburx=568,bbury=540}
\vspace{-14.5mm}
\caption{Comparison of quenched data and fit (3) with the
dynamical results. The star indicates the experimental value.}
\end{figure}

\vspace*{-2\baselineskip} In Figure~4 we plotted our quenched results
together with the dynamical ones.  The line represents the fit~(3).
The fit range is much larger than the range of this plot which only
includes the six smallest masses (open symbols).  Towards the chiral
limit we see that the dynamical data (filled symbols) lie
systematically below the quenched data.  The dynamical results are
consistent with the quenched fit.

The corresponding picture for $\mN$ is shown in Figure~5.  The open
symbols represent quenched results and the line is a phenomenological
fit like (3) where the curvature $b_3$ is again small.  The filled
symbols are dynamical results.  The agreement between quenched and
dynamical results is more pronounced than for $\mV$.

\subsection{Conclusions}

We observe that the results with non-perturbatively $O(a)$ improved
dynamical fermions scale well.  Although we are not able to do a
continuum extrapolation yet, we see no large discretization
errors.  In the considered range of sea quark masses we
found no clear evidence for unquenching effects in comparison with
previous quenched results.

However, our results seem to indicate that the

\pagebreak

\begin{figure}[t]
\epsfig{file=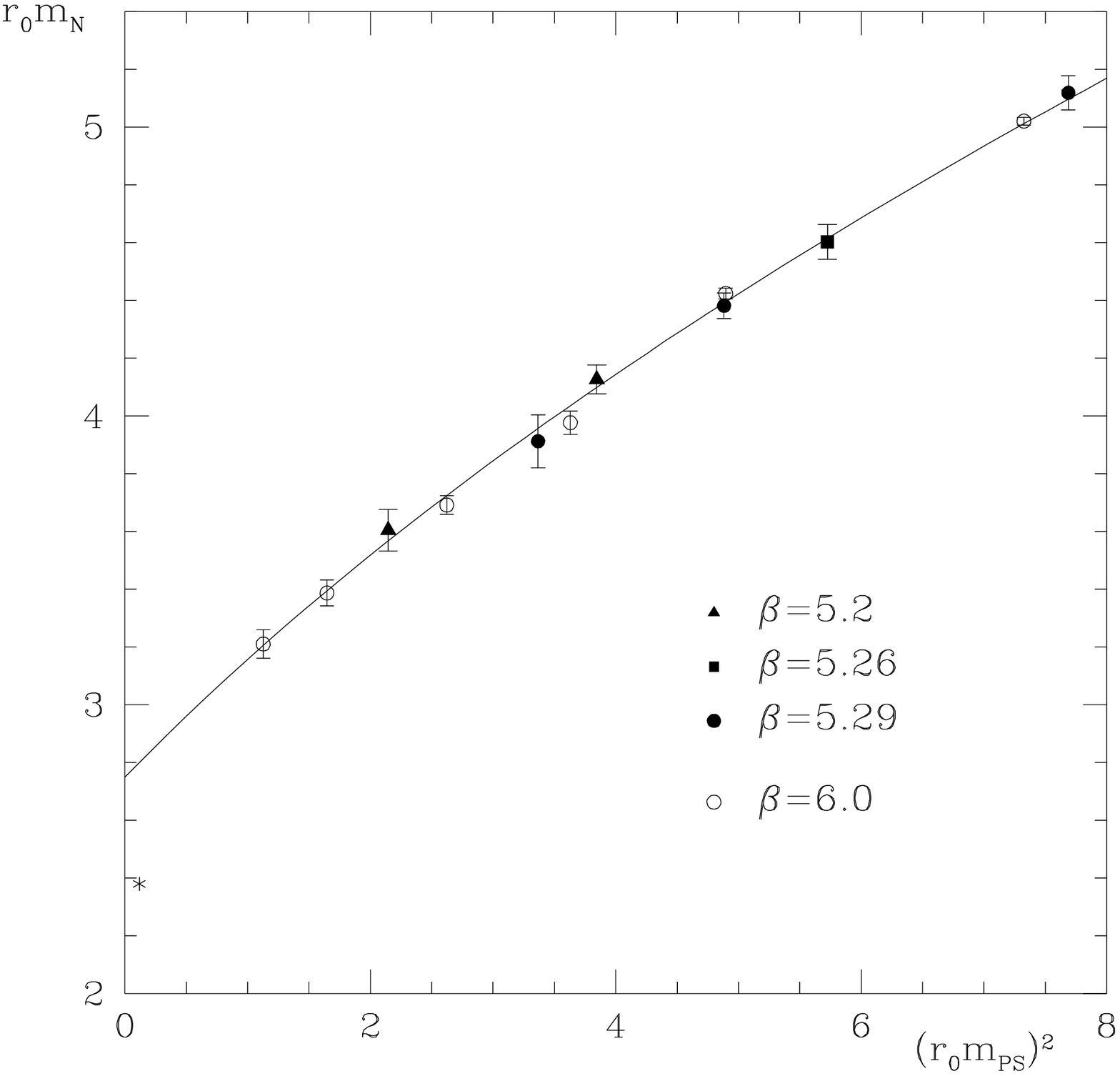,width=7.5cm,bbllx=0,bblly=0,bburx=568,bbury=540}
\vspace{-14.5mm}
\caption{Comparison of $\mN$ quenched data and fit like in
Figure~4 with the dynamical results. The star indicates the experimental
value.}
\end{figure}

\vspace*{-1.8\baselineskip}
\noindent
vector meson mass $\mV$ approaches the chiral limit 
in a different
way.  This may be another hint, that quenched artifacts are visible
in the quenched data.

\section*{ACKNOWLEDGEMENTS}

The computations were done on the Hitachi SR8000-F1 at LRZ (Munich),
on the Cray T3Es at EPCC (Edinburgh), NIC (J\"ulich), (ZIB) Berlin and
on the APE/Quadrics at DESY Zeuthen.  We thank these institutions for
support.  UKQCD acknowledges PPARC grants GR/L22744 and
PPA/G/S/1998/000777.

\pagebreak
\end{document}